\documentstyle[aps,prl,epsfig]{revtex}
\input epsf

\newcommand{\eq} [1]{equation~(\ref{eq:#1})}
\newcommand{\fig}[1]{Figure~(\ref{fig:#1})} 
\newcommand{\tbl}[1]{Table~(\ref{tbl:#1})} 
\newcommand{\figtunneling}{
\begin{figure}[h]
\vspace*{-1.2cm}
\centerline{\epsfig{figure=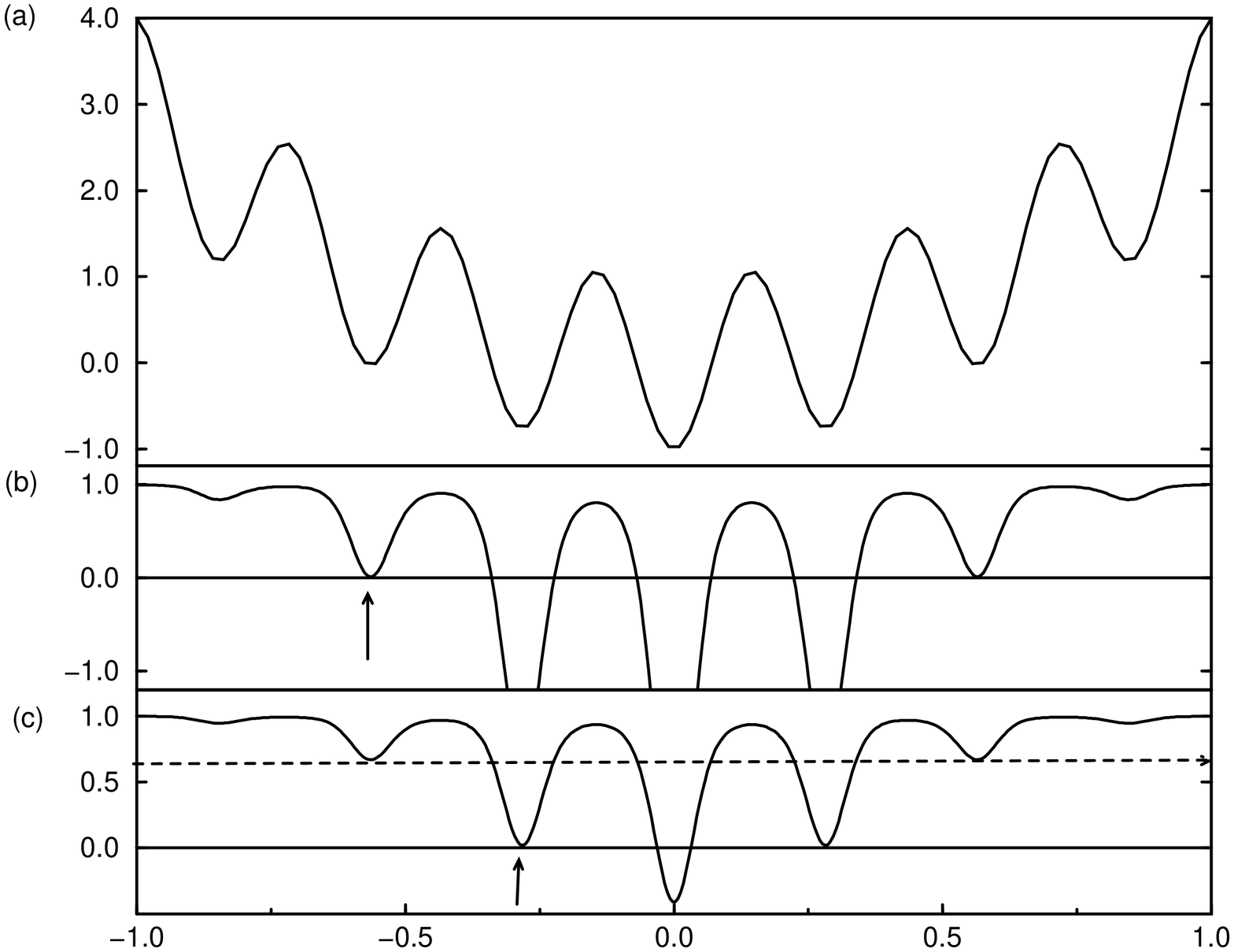,height=9cm,angle=-90}}
\caption{(a) Schematic one-dimensional PES  and (b) 
STUN effective potential, where the minimum indicated by the arrow is
the best minimum found so far. All wells that lie above the best
minimum found are suppressed. If the dynamical process can escape the
well around the current ground-state estimate it will not be trapped
by local minima that are higher in energy. Wells with deeper minima
are preserved and enhanced. (c) After the next minimum to the right
has been located, wells that were still pronounced in (b) are also
suppressed, now only the wells around the improved ground-state
estimate and the true ground state are pronounced. Once the true
ground state has been found (not shown) all other wells have been
suppressed and will no longer trap the dynamical process. The dotted
line in (c) illustrates an energy threshold $0 < f_t < 1$ to classify
the nature of the dynamics. In order to conduct a successful search
the dynamical process must explore both paths confined to the vicinity
of the present well ($f_{\rm STUN} < f_t$) and paths that escape the
well by tunneling the barrier ($f_{\rm STUN} > f_t$). Adjusting the
temperature to maintain a particular average effective energy balances
the tunneling and the local-search phases of the algorithm.}
\label{fig:transform}
\end{figure}
}

\newcommand{\figglass}{
\begin{figure}[h]
\vspace*{-1.5cm}
\centerline{\epsfig{figure=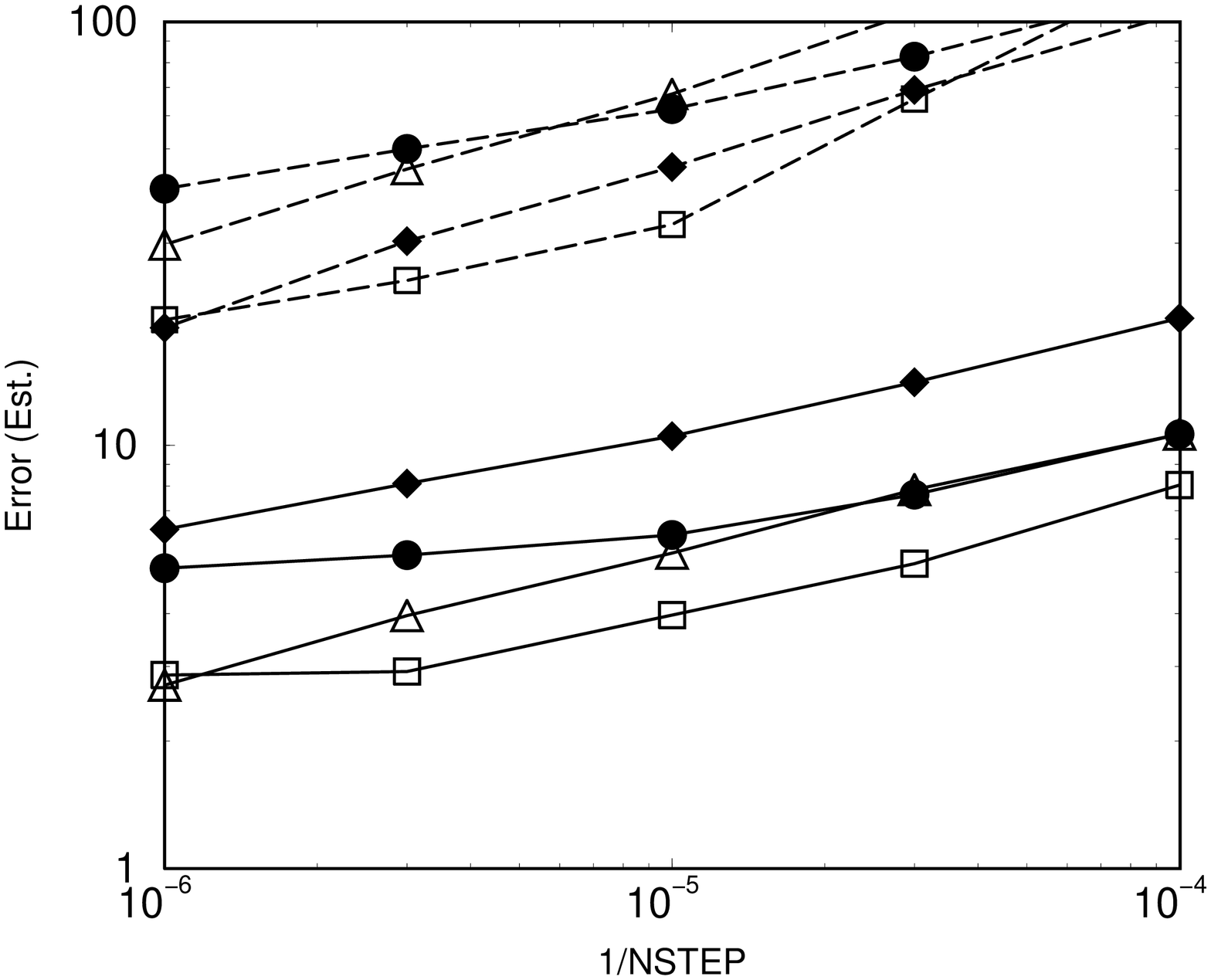,height=9cm,angle=-90}}
\caption{Average estimated error for the ground-state estimates of the 
Coulomb glass using SA (circles), STUN(squares), ST(triangles) and
PT(diamonds) for N=100 (full lines in lower part) and N=500 (dashed
lines in upper part).
\label{fig:lrglass}}
\end{figure}
}

\newcommand{\tabletsp}{
\begin{table}
\twocolumn[\hsize\textwidth\columnwidth\hsize\csname@twocolumnfalse%
\endcsname
\caption{Estimates for the optimal path-length for the 
traveling salesman problem with $N=20,50$ and $100$ sites using either
only local (left side) or global (right side) moves as described in
the text. For global moves both SA and STUN are equally efficient to
obtain low-energy paths. Using only local moves the existence of
barriers hampers the progress of SA.  As a result SA becomes less
efficient than STUN. By virtue of its temperature exchange mechanism
PT also allows the random walk to cross the barriers, but is less
efficient than STUN. The effort is given in thousands of steps, note
that the evaluation of a local move is much less costly than that of a
global move. The path-length indicate the average optimal energy for
20 runs and the best energy found. \label{tbl:tsp}}
\begin{center}
\begin{tabular}{|r|r|rrr||rr|}
     &            & \multicolumn{3}{c||}{Local Moves}  & \multicolumn{2}{c|}{Global Moves} \\ 
\hline
N    &Effort &       SA        &       PT       &      STUN      & SA    &        STUN  \\
\hline
20   &   50  &   4.85 /  3.55  &  4.35 /  3.55  &  3.60 / 3.55 &   3.94 / 3.61  &  3.55 / 3.55\\
20   &  100  &   4.52 /  3.58  &  4.02 /  3.55  &  3.62 / 3.55 &   3.93 / 3.55  &  \\
20   &  500  &   4.08 /  3.55  &  3.57 /  3.55  &  3.55 / 3.55 &   3.82 / 3.56  &  \\
20   & 1000  &   4.08 /  3.55  &  3.55 /  3.55  &	       &	        &\\
20   & 5000  &   3.75 /  3.55  &	        &	       &	        &\\
\hline       	      		        	         	        	         
50   &  100  &   12.5 / 10.61  & 13.72 / 12.58  & 11.06 / 9.39 &   5.74 / 5.65  &  5.72 / 5.65\\
50   &  500  &   11.0 /  8.68  & 11.55 / 10.65  &  8.32 / 5.83 &   5.70 / 5.65  &  5.67 / 5.65\\
50   & 1000  &   11.0 /  8.84  & 10.70 /  9.82  &  7.75 / 5.78 &   5.68 / 5.65  &  5.67 / 5.65\\
50   & 5000  &   9.84 /  8.10  &  8.99 /  7.89  &  7.16 / 5.78 &   5.66 / 5.65  &  5.65 / 5.65\\
50   &10000  &   9.87 /  8.31  &                &  6.70 / 5.72 &   5.66 / 5.65  & \\
\hline      	      		        	         	        	         
100  &  100  &                 &                &              &   8.42 / 8.11  &  8.40 / 8.01 \\ 
100  &  500  &                 &                &              &   8.18 / 8.01  &  8.18 / 7.97 \\
100  & 1000  &                 &                &              &   8.08 / 7.94  &  8.03 / 7.95 \\
100  & 5000  &                 &                &              &   8.01 / 7.94  &  8.01 / 7.96 \\
\end{tabular}
\end{center}
]
\end{table}
}

\newcommand{\tablelabs}{
\begin{table}
\caption{Average and best ground state estimates for LABS for the $N=49$
and $N=101$ using SA and STUN on the locally minimized PES described
in the text. SA now systematically approaches the estimated
ground-state energy, but STUN is about two orders of magnitude more
efficient. The effort is given in thousands of steps, each step
consists of a multi-spin flip followed by a local minimization.\label{tbl:labs}}

\begin{tabular}{|r|rr|}
Effort    &       SA     &          SC \\
\hline
      & \multicolumn{2}{c|}{N = 49} \\
\hline
  10  &  212.48 / 176  &   185.12 / 136 \\
  50  &  196.64 / 164  &   168.72 / 136 \\
 100  &  191.68 / 144  &   161.60 / 136 \\
 500  &  177.68 / 136  &   151.76 / 136 \\
1000  &  175.52 / 136  &   139.44 / 136 \\
\hline		        
      & \multicolumn{2}{c|}{N = 101} \\
\hline
  10  &  987.44 / 914  &   918.08 / 810 \\
  50  &  946.44 / 854  &   880.08 / 790 \\ 
 100  &  927.84 / 846  &   865.76 / 766 \\    
 500  &  894.32 / 822  &    \\
1000  &  891.68 / 818  &    \\
\end{tabular}
\end{table}
}

\begin{document}
\draft
\twocolumn[\hsize\textwidth\columnwidth\hsize\csname@twocolumnfalse%
\endcsname

\title{A Stochastic Tunneling Approach for Global Minimization of
Complex Potential Energy Landscapes}
\author{W. Wenzel and K. Hamacher}
\address{Institut f\"ur Physik, Universit\"at Dortmund, \\D-44221
Dortmund, Germany}
\maketitle
\begin{abstract} 
We investigate a novel stochastic technique for the global
optimization of complex potential energy surfaces (PES) that avoids
the freezing problem of simulated annealing by allowing the dynamical
process to tunnel energetically inaccessible regions of the PES by way
of a dynamically adjusted nonlinear transformation of the original
PES. We demonstrate the success of this approach, which is
characterized by a single adjustable parameter, for three generic hard
minimization problems.\\

\vspace{1em}
\end{abstract}

\vskip 0.3 in
]

The development of methods that efficiently determine the global
minima of complex and rugged energy landscapes remains a challenging
problem with applications in many scientific and technological
areas. In particular for NP-hard\cite{compcomplex,compcomplex2}
problems stochastic methods offer an acceptable compromise between the
reliability of the method and its computational cost. Branch-and-bound
techniques\cite{barahona88} offer stringent error estimates but scale
exponentially in their computational effort. In many stochastic
approaches the  computational cost to determine the global
minimum with a given probability grows only as a power-law with
the number of variables~\cite{hamacher98}.

In such techniques the global minimization is performed through the
simulation of a dynamical process for a ``particle'' on the
multi-dimensional potential energy surface. Widely used is the
simulated annealing (SA) technique~\cite{kirkpatrick83} where the PES
is explored in a series of Monte-Carlo simulations at successively
decreasing temperatures. Its success depends often strongly on the
choice of the cooling schedule, yet even the simplest geometric
cooling schedule is characterized by three parameters (starting
temperature, cooling rate and number of cooling steps) that must be
optimized to obtain adequate results. For many difficult problems with
rugged energy landscapes SA suffers from the notorious ``freezing''
problem, because the escape rate from local minima diverges with
decreasing temperature. To ameliorate this problem many variants of
the original algorithm\cite{berg91,lyubartsev92,hansmann97}
have been proposed. Unfortunately these proposals often increase
the number of parameters even further, which complicates their
application for practical problems.

In this letter we investigate the stochastic tunneling method, a
generic physically motivated generalization of SA. This approach
circumvents the ``freezing'' problem, while reducing the number of
problem dependent parameters to one. In this investigation we demonstrate the
success of this approach for three hard minimization problems: the
Coulomb spin-glass (CSG), the traveling salesman problem (TSP) and the
determination of low autocorrelation binary sequences(LABS) in
comparison with other techniques.

{\em Method:} The freezing problem in stochastic minimization methods
arises when the energy difference between ``adjacent'' local minima on
the PES is much smaller than the energy of intervening transition
states separating them. As an example consider the 
dynamics on the model potential in \fig{transform}(a). At
high tem\-peratures a particle can still cross the barriers, but not
differentiate between the wells. As the temperature drops, the
particle will get eventually trapped with almost equal probability in
any of the wells, failing to resolve the energy difference between
them. The physical idea behind the stochastic tunneling method is to
allow the particle to ``tunnel''\cite{barhen97} forbidden regions
of the PES, once it has been determined that they are irrelevant
for the low-energy properties of the problem. This can be accomplished
by applying a non-linear transformation to the PES:
\begin{equation}
f_{\rm STUN}(x) = 1 - \exp\left[ -\gamma (f(x) - f_0)\right]
\label{eq:transform}
\end{equation}
where $f_0$ is the lowest minimum encountered by the dynamical process
so far (see \fig{transform}(b) + (c))\cite{comment}. The effective potential
preserves the locations of all minima, but maps the entire energy
space from $f_0$ to the maximum of the potential onto the interval
$[0,1]$. At a given finite temperature of O(1), the dynamical process
can therefore pass through energy barriers of arbitrary height, while
the low energy-region is still resolved well. The degree of steepness
of the cutoff of the high-energy regions is controlled by the
tunneling parameter $\gamma > 0$. Continously adjusting the reference
energy $f_0$ to the best energy found so far, successively eliminates
irrelevant features of the PES that would  trap the dynamical
\nobreak process.

To illustrate the physical content of the transformation we  
consider a Monte-Carlo (MC) process at some fixed inverse temperature
$\beta$ on the STUN-PES. A MC-step from $x_1$ to $x_2$ with 
$\Delta = f(x_2) - f(x_1)$ is accepted with probability 
$ \tilde{w}_{1\rightarrow 2} = \exp\left[-\beta(f_{\rm STUN}(x_2) - f_{\rm
STUN}(x_1))\right]$. In the limit  $\gamma\Delta \ll 1$ this reduces to 
$\tilde{w}_{1\rightarrow 2} \approx \exp(-\tilde{\beta}\Delta)$  
with an effective, energy dependent temperature $\tilde{\beta} = \beta
\gamma e^{\gamma ( f_0 - f(x_1))} \le
\beta\gamma.$ 
The dynamical process on the STUN potential energy
surface with fixed temperature can thus be interpreted as an MC process
with an energy dependent temperature on the original PES. In the
latter process the temperature rises rapidly when the local energy
is larger than $f_0$ and the particle diffuses (or tunnels) freely through
potential barriers of arbitrary height. As better and better minima
are found, ever larger portions of the high-energy part of the PES are
flattened out. In analogy to the SA approach this behavior can be
interpreted as a self-adjusting cooling schedule that is optimized as
the simulation proceeds. 

Since the transformation in \eq{transform} is bounded, it is possible
to further simplify the method: 
On the fixed energy-scale of the effective potential one can
distinguish between phases corresponding to a local search and
``tunneling'' phases by comparing $f_{\rm STUN}$ with some fixed,
problem independent predefined threshold $f_{\rm t}$ (see
Fig. 1(c)). For the success of the method it is essential that the
minimization process spends some time tunne-

\figtunneling

\figglass

\noindent ling and some time
searching at any stage of the minimization process. We therefore
adjust the parameter $\beta$ accordingly during the simulation: If a
short-time moving average of $f_{\rm STUN}$ exceeds the threshold
$f_{\rm thresh} \approx 0.03$, $\beta$ is reduced by some fixed
factor, otherwise it is increased. Following this prescription the
method is characterized by the single problem-dependent
parameter ($\gamma$). 
%
%  Explanation
%
%While the transformation chosen \eq{transform} is only one of many
%possible choices transformation, we believe that there are a number of
%features that constrain its construction: (i) The transformation must
%be strongly nonlinear in the high-energy regime, as only such a
%transformation will lead to a nearly constant effective PES for high
%energies and true ``tunneling''. (ii) There must be a parameter that
%modulates the degree of compression ($\gamma$), since the ratio the
%energy differences of adjacent local minima to the transition state
%energy separating them varies from problem to problem. (iii) Requiring
%an essentially flat PES at high energy (for typically unbounded PES)
%requires a transformation that maps the interval $[f_0,\infty]$ onto
%some finite interval, which can be chosen as $[0,1]$ without loss of
%generality.

{\em Applications:} In order to test the performance of this algorithm
we have investigated three families of complicated NP-hard
minimization problems. For each problem we have determined either the
exact ground-state energy or a good estimate thereof. We computed
the average error of the various optimization methods as a function of
the computational effort to determine the computational effort
required to reach a prescribed accuracy. 
For the applications presented
here we have fixed the functional form of the transformation and the
``cooling schedule'' for $\beta$ in order to demonstrate that these
choices are sufficient to obtain adequate results. Obviously this does
not guarantee that these choices are optimal.

(CSG) The determination of low-energy configurations of glassy PES is
a notoriously difficult problem. We have verified by direct comparison
that the method converges quickly to the exact ground
states~\cite{simone95} for two-dimensional short-range Ising
spin-glasses of linear dimension $10$ to $30$ with either discrete or
Gaussian distributions of the coupling parameters. Next we turned to
the more demanding problem of the Coulomb spin-glass, where classical
charges $\{s_i\}$ with $s_i = \pm 1$ are placed on fixed randomly
chosen locations within the unit  cube. The energy of the system
\begin{equation}
 E(\{s_i\}) = \sum_{ij}^{N} \frac{s_i\ s_j}{|\vec{r}_i -\vec{r}_j|},
\end{equation}

\tabletsp
\noindent is minimized as a function of the distribution of the $\{s_i\}$.  

The results of grand-canonical simulations for ten replicas of $N=100$
and $N=500$ charges are shown in Figure 2. We first conducted twenty
very long STUN runs for each replica to determine upper bounds for the
true ground-state energy. For the same charge distributions we then
averaged the error of STUN, SA, parallel tempering (PT)~\cite{hansmann97} 
and simulated tempering(ST)~\cite{lyubartsev92} for twenty runs per replica as
function  of the numerical effort.  We found that the average STUN
energy converged in $10^6$ MC-steps to within 1\% of the estimated
true ground-state energy.  Over two decades of the numerical effort we
found a consistent significant advantage of the proposed method over
the SA approach. Fitting the curves in the figure with a power-law
dependence we estimate that STUN is two orders of magnitude more
efficient than SA.

We found no consistent ranking of ST and PT relative to SA for the two
system sizes considered. Both methods offer alternative routes to
overcome the freezing problem in SA. In PT the configurations of
concurrent simulations at a variety of temperatures are occasionally
exchanged. In ST only a single simulation is undertaken, but its
temperature is considered to be a dynamical variable. Temperature and
configuration are distributed according to: $ p(s,T) = e^{-E(s)/T -
g(T)} $ and the weights $g(T)$ are optimized for a discretized
temperature distribution, such that all temperatures are visited with
equal probability. In both methods, a configuration can escape a local
minimum when the instantaneous temperature is increased.  The choice
of the temperature set (along with values for $g(T)$) is system
dependent and must be optimized much like the annealing schedule in
SA. In accordance with other studies our results indicate that ST performs
significantly better than SA for long simulation times. PT was
successful only for the larger system (N=500), where it reached the
same accuracy as STUN for $10^6$ steps. STUN converged faster than any
of the competing methods, but showed a tendency to level off at high
accuracy. In the limit of large computational its accuracy was matched
by ST for N=100 and PT for N=500.

(TSP) The traveling salesman problem is another ubiquitous NP-hard
minimization problem\cite{press95,dittes96}. We have investigated the
problem in its simplest incarnation: i.e. as \noindent a minimization
of the euclidian distance along a closed path of N cities. Using
long-range updates, i.e. the reversal and exchange of paths of
arbitrarily length, we found that both SA and STUN perform about
equally well and reach the global optimum for $N=20,50$ and $100$ very
quickly (see right side of~\tbl{tsp}).

Nevertheless it is instructive to analyze this model somewhat further
as it provides insight into the interplay of move-construction and
complexity of the minimization problem.  The unconstrained TSP is a
rare instance among NP-hard minimization problems, where it is
possible to construct efficient ``long-range'' hops on the PES. In
most practical applications of minimization problems related to the
TSP, the construction of global moves is severely complicated by the
existence of ``hard constraints'' on the routes taken. For such
problems, as well as the other examples reported here, the alteration
of just a few variables of the configurations leads to unacceptably
high energies in almost all cases. As a result, the construction of
global moves is not an efficient way to facilitate the escape from
local minima.  When only local moves, i.e. transpositions of two
adjacent cities, are considered high barriers that were
circumvented in the presence of global moves hamper the progress of
SA. The results on the left side of \tbl{tsp} demonstrate that in this
scenario SA performs significantly worse than STUN. 

(LABS) Finally we turn to the construction of low-autocorrelation
binary sequences\cite{bernasconi87,dittes96}. The model can be cast as
a ground-state problem for a one-dimensional classical spin-1/2 chain
with long-range four-point interactions
\begin{equation}
 E = \frac{1}{N} \sum_{k = 1}^{N-1} \left[ \sum_{j=1}^{N-k} s_j \ s_{j+k}\right]^2
\end{equation}
and is one of the hardest discrete minimization problems
known\cite{krauth95}. Even highly sophisticated and specialized
optimization algorithms\cite{dittes96} have failed to find
configurations anywhere near (within 20\%) the ground-state energy
that can be extrapolated from exact enumeration studies for small
systems ($N < 50$)\cite{golay82,beenker85}. The reason for this
difficulty has been attributed to the ``golf-course'' character of the
energy landscape and there is convincing evidence that SA will fail to
converge to the ground-state energy even in the limit of adiabatic
cooling\cite{bernasconi87}. The situation is significantly improved if
the original potential energy surface is replaced by a piecewise
constant energy surface that is obtained by a local minimization of
the original PES at each point\cite{scheraga91}. Obviously the latter
surface preserves all ground-state configurations and energies of the
original PES, but eliminates many ``plateaus'' of the ``golf-course''
landscape.  Using the modified energy surface we are able to compare
SA to STUN, since SA can now determine the ground
state energy of medium 

\tablelabs

\noindent 
size systems (N=49) with a large, but finite
computational effort. Table II summarizes the results for the average
error of 20 SA and STUN runs for system sizes N=49 and N=101 as a
function of the computational effort. In direct comparison we find
that STUN is two orders of magnitude more efficient than SA. Both
methods are at least a dozen orders of magnitude more efficient than
SA on the original PES.

{\em Discussion:} Using three NP-hard minimization problems with
high-barriers separating local minima we have demonstrated that the
stochastic tunneling approach offers a reliable, generic and efficient
route for the determination of low-energy configurations. One chief
advantage of the method lies in the fact that only a single
parameter must be adjusted to adapt it for a specific
problem. 
%Heuristically we observed that optimal values for this
%parameter are of the order $\langle \gamma \Delta E \rangle
%\sim O(10^{-3})$, where $\Delta E$ is a typical energy-change of an
%allowed move. 
One of the drawbacks of STUN is that in contrast to e.g. PT, no
thermodynamic expectation values for the system can be obtained from
the simulation. Secondly, because the non-linear transformation will
map any unbounded PES onto an interval bounded from above, the
dynamical process in STUN will experience ``tunneling'' phases at any
finite temperature. For PES that do not contain high barriers, or in
the presence of efficient global moves that circumvent such barriers,
STUN may therefore be less efficient than competing methods. In many
realistic optimization problems where the construction of global moves
is exceedingly difficult or very expensive the tunneling
approach can ameliorate the difficulties associated with
the existence of high energy barriers that separate local minima of
the PES.

We gratefully acknowledge stimulating discussions with C. Gros and
U. Hansmann.

%\bibliographystyle{unsrt}     
%\bibliography{optimization}

\begin{thebibliography}{10}
\bibitem{compcomplex}
C.~H. Papadimitriou.
\newblock {\em Computational Complexity}.
\newblock Addison-Wesley, Reading, Massachusetts, 1994.

\bibitem{compcomplex2}
M.~R. Garey and D.~S. Johnson.
\newblock {\em Computers and Intractability - A Guide to the Theory of
  NP-Completeness}.
\newblock Freeman and Company, New York, 1979.

\bibitem{barahona88}
F.~Barahona, M.~Gr\"otschel, M.~J\"unger, and G.~Reinelt.
\newblock {\em Operations Research}, 36:493, 1988.

\bibitem{hamacher98}
K.~Hamacher and W.~Wenzel.
\newblock {\em PRE}, 58:938, 1999.

\bibitem{kirkpatrick83}
S.~Kirkpatrick, C.D. Gelatt, and M.P. Vecchi.
\newblock {\em Science}, 220:671--680, 1983.

\bibitem{berg91}
B.~A. Berg and T.~Neuhaus.
\newblock {\em Phys. Letters}, B267:249, 1991;
J.~Lee. \newblock {\em PRL}, 71:211, 1993;
B.~Hesselbo and R.~B. Stinchcombe.
\newblock {\em PRL}, 2151:1995, 74.

\bibitem{lyubartsev92}
A.~P. Lyubartsev, A.A. Martinovski, S.~V. Shevkunov, and P.N.
  Vorontsov-Velyaminov.
\newblock {\em JCP}, 96:1776, 1992;
E.~Marinari and G.~Parisi.
\newblock {\em EPL}, 451:1992, 19.;

\bibitem{hansmann97}
U.~H.~E. Hansmann. \newblock {\em CPL}, 281:140, 1997.

\bibitem{comment} 
Considerations motivating the choice of this functional and
constraints on its construction will be dicsussed in a forthcoming
publication.

\bibitem{barhen97}
J. Barhen, V. Protoposecu, and D. Reister.
\newblock {\em Science}, 276:1094--1097, 1997.

 
\bibitem{simone95}
C.Simone, M.~Diehl, { M.~J{\"u}nger}, P.~Mutzel, and G.~Reinelt.
{\em J. Stat. Phys.}, 80:487, 1995.
http://www.informatik.uni-koeln.de/ls\_juenger/projects/sgs.html.

\bibitem{press95}
W. H.~Press et~al.
\newblock {\em Numerical Recipies in C}.
\newblock Cambridge University Press, Cambridge, 1995.


\bibitem{bernasconi87}
J.~Bernasconi.
\newblock {\em J. Physique}, 48:559, 1987.

\bibitem{dittes96}
F.-M. Dittes.
\newblock {\em Phys. Rev. Lett.}, 76(25):4651--4655, 1996.

\bibitem{krauth95}
W.~Krauth and M.~Mezard.
\newblock {\em Z. Phys. B}, 97:127, 1995.

\bibitem{golay82}
M.~J.~E. Golay.
\newblock {\em IEEE Trans. Inform. Theory}, 28:543, 1982.

\bibitem{beenker85}
G.~F.~M. Beenker, T.~Claasen, and P.~W.~C. Hermes.
\newblock {\em Philips J. Res.}, 40:289, 1985.

\bibitem{scheraga91}
A. Nayeem, J. Vila, and H.~A. Scheraga.
\newblock {\em J. Comp. Chem.}, 12(5):594--605, 1991.

\end{thebibliography}

\end{document}